\documentclass[useAMS,amssymb,epsfig,usegraphicx,twocolumn,usenatbib]{mn2e}
\def\aj{AJ}% Astronomical Journal
\def\araa{ARA\&A}
\def\apj{ApJ}% Astrophysical Journal
\def\apjl{ApJ}% Astrophysical Journal, Letters
\def\aap{A\&A}% Astronomy and Astrophysics
\def\mnras{MNRAS}% Monthly Notices of the RAS
\def\na{New Astronomy} % New Astronomy
\def\nat{Nature}% Nature\def\pasj{PASJ} % Publications of Astronomical Society Japan
\def\aaps{A\&AS} 

\usepackage[usenames,dvips]{color}

\newif\ifAMStwofonts

\makeatother

\newif\ifAMStwofonts

\title[Disc heating]
{Disc heating: possible link between weak bars and superthin galaxies}
\author[Saha]
{Kanak Saha $^{}$\thanks{E-mail:kanak@iucaa.ernet.in} \\
Inter University Centre for Astronomy and Astrophysics, Post Bag 4, Ganeshkhind, Pune 411007, India}
 
\begin{document}

\date{Accepted xxxx Month xx. Received xxxx Month xx; in original form
\today} 
\pagerange{\pageref{firstpage}--\pageref{lastpage}} \pubyear{2014}
\maketitle

\label{firstpage}

\begin{abstract}
The extreme flatness of stellar discs in superthin galaxies is 
puzzling and the apparent dearth of these objects in cosmological simulation
poses challenging problem to the standard cold dark matter paradigm. Irrespective of 
mergers or accretion that a galaxy might be going through, stars are 
heated as they get older while they interact with the spirals and bars 
which are ubiquitous in disc galaxies -- leading to a puffed up stellar 
disc. It remains unclear how superthin galaxies maintain their
thinness through the cosmic evolution.

We follow the internal evolution of a sample of 16 initially extremely 
thin stellar discs using collisionless N-body simulation. All of these discs 
eventually form a bar in their central region. Depending on the initial condition, 
some of these stellar discs readily form strong bars while others grow weak bars 
over secular evolution time scale. We show that galaxies with strong bars heat 
the stars very efficiently, eventually making their stellar discs thicker. On 
the other hand, stars are heated very slowly by weak bars -- as a result, galaxies 
hosting weak bars are able to maintain their thinness over several billion years, 
if left isolated. We suggest that some of the superthin galaxies might as well be forming 
weak bars and thereby prevent any strong vertical heating which in turn helps maintaining 
their thinness during the course of secular evolution. 

\end{abstract}


\section{Introduction}
\label{sec:intro}
Galaxy formation under the $\Lambda$CDM cosmology produces too few thin, disc-dominated
late-type galaxies with little or essentially no bulge \citep{MoMaoWhite1998,DOnghiaBur2004, 
Mayeretal2008}. On the other hand, up-to-date observational surveys show abundance of flat 
galaxies \citep{Karachentsevetal1999, Matthews2000b, kautschetal2006} in the local universe. 
The apparent dearth of very thin disc galaxies in cosmological simulation has put up a 
challenge against the standard theory of CDM cosmology. 

Superthin galaxies (hereafter SGs) are the extreme case of these flat edge-on galaxies 
with their axial ratios (defined as the ratio of scale height to the scale length) generally below
$0.1$. Due to the lack of strong morphological features SGs belong to the 
late-type in the Hubble classification scheme. In few cases, the thin discs of SGs 
are seen to be warped e.g., UGC 7170, UGC 3697 (in which case, the axial ratios are $\sim 0.1$) 
\citep{Karachentsevetal1999} which could be considered as the limiting case of a 
superthin galaxy. SGs do have other unique properties of the late-type galaxies e.g., 
low surface brightness (LSB), high atomic hydrogen gas fractions, low metallicities 
\citep{vanderHulstetal1993}. The extreme thinness of the stellar discs suggest that 
SGs somehow prevent strong heating in the vertical direction. It is puzzling how 
SGs maintain their superthinness over several billion years since they are formed.  
 
Observations point out that SGs, like many LSB galaxies, lack environmental influences
\citep{Rosenbaumetal2009,Galazetal2011} 
which could give
rise to tidal heating, heating due to satellite infall or mergers. The discs of 
superthin galaxies e.g., UGC 7321, IC2233 appear to be featureless, smooth with 
no visible signs of interactions such as tidal streams or other irregularities 
\citep{Matthewsetal1999, Matthews2008}. In contrast, the resulting stellar discs 
of CDM simulations are often smaller and thicker \citep{Kazantzidisetal2008}. This 
indicates that the heating of stellar discs of SGs are unlikely due to satellite 
infall, merger or due to massive subhalos (as it is in the CDM simulations). 
Since external mechanisms appear to be unimportant in heating stars of SGs, one must 
rely upon various internal sources. Indeed, as a galaxy evolves, internal heating 
of stars in the disc is unavoidable. In the case of our Milky Way, it is evident 
from the Hipparcos data \citep{Binneyetal2000}. Recently, it has been shown by \cite{Sahaetal2010} 
that a strong bar could efficiently heat stars in the vertical direction through
physical processes like chaotic diffusion suggested by \cite{Pfenniger1985}. 
In fact, the possibility that superthin galaxies might be harbouring thin bars e.g., 
in UGC 7321 \citep{ Uson2003, Pohlenetal2003} could not be ruled out unambiguously. 
Bars are ubiquitous in disc galaxies \citep{Eskridgeetal2000,Barazzaetal2008} suggesting
that they are formed spontaneously and perhaps survive through the cosmic evolution
\citep[see][for a recent review]{Sellwood2013}. However, depending on the initial 
physical condition of a stellar disc and the distribution of dark matter, 
bars could grow to be stronger over a few rotation time scales or remain weak 
over several billion years \citep{Sahaetal2010}. The analysis of \cite{Sahaetal2010} 
further indicates that weak bars are not efficient in vertical heating of stars --
implying that if an initially superthin galaxy were to prevent strong vertical heating, 
it could possibly do so by growing a weak bar at the most during the course of 
secular evolution.  

The main focus of the current paper is to follow the evolution of such superthin 
galaxies embedded in various dark matter halo configuration and find out how many of these
initial superthin galaxies are able to maintain their thinness such that they are still
classified as superthin. We use collisionless
N-body simulations to study the evolution of $16$ initially superthin galaxies
in isolation. Our study suggest that weak bars are perhaps the maximal 
non-axisymmetric features that a self-consistent superthin galaxy might 
be able to support in order to maintain their thinness over several Gyr.
 
The paper is organized in the following way. Section~\ref{sec:model} describes 
the models of superthin galaxies and N-body simulation. Section~\ref{sec:heating} 
summarizes plausible sources of heating relevant for superthin galaxies. The 
detailed evolution of superthin galaxies is presented in section~\ref{sec:evolution}. 
Section~\ref{sec:discuss} contains the discussion and primary conclusions from this work.  

\section{Initial galaxy models}
\label{sec:model}
The properties of stellar discs of superthin galaxies resemble many of the key 
properties of the late-type galaxies, especially that of the low surface 
brightness (LSB) galaxies \citep{McGaughetal1995,Matthews2000}. Like LSB galaxies,
 the dynamics in superthin galaxies are thought to be primarily governed by their 
surrounding dark matter halos. The dark matter dominance in LSB galaxies is evident from the 
decomposition of rotation curves \citep{deBloketal2001} and in the case of 
superthin galaxy UGC~7321 \citep{Uson2003} from the combined modelling of the 
rotation curves and HI scale~height \citep{banerjeeetal2010}. The SGs are known 
to have poor star formation activities despite having higher neutral hydrogen gas 
fraction as in high surface brightness galaxies \citep{vanderHulstetal1993,Boissieretal2008}. The very 
low star formation rates 
and the apparent absence of strong two-armed spirals and/or strong bars suggest 
that the stellar discs in superthin galaxies are probably dynamically hot --
indicative of a high value of Toomre's stability parameter (Q) \citep[see,][]{BT1987}. 
We utilize these general properties as guidelines to construct initial N-body models of extremely 
thin galaxies some of which are presumably progenitors of present day superthin galaxies. 

\begin{table}
\caption[ ]{Basic parameters for the model galaxies and 
bar strength (defined in section~\ref{sec:sbar}) computed at the end of $5$~Gyr.}
\begin{flushleft}
\begin{tabular}{ccccccccc}  \hline\hline 
Galaxy    & Q & $\frac{\sigma_z}{\sigma_r}$ & $M_d$ & $\frac{M_h}{M_d}$ & $R_c/R_d$ & $A_2/A_0$ &\\% comments & \\
models &     &  &($\times 10^{10} M_{\odot}$) &     &      &   &   \\
\hline
\hline
RHG041 & 1.97 & 0.27 & 2.00 & 4.73 & 1.05 & 0.165 \\
RHG040 & 2.23 & 0.34 & 1.86 & 7.67 & 1.70 & 0.33 \\
RHG053 & 2.31 & 0.30 & 1.05 & 11.93 & 0.90 &0.082 \\
RHG052 & 2.33 & 0.62 & 1.43 & 14.47 & 2.07 & 0.37 \\ 
RHG102 & 2.56 & 0.37 & 1.28 & 12.94 & 0.80 &0.15 \\
RHG114 & 2.97 & 0.26 & 1.59 & 10.17 & 1.25 &0.13 \\
RHG057 & 2.98 & 0.39 & 0.86 & 22.53 & 1.03 &0.21 \\
RHG034 & 3.15 & 0.25 & 1.92 & 6.50 & 0.81 &0.154 \\
\hline
RCG051 & 1.21 & 0.41 & 4.64 & 6.75 & 1.88 & 0.524 \\
RCG051A &1.21 & 0.61 & 4.75 & 6.52 & 1.98 & 0.508 \\
RCG049 & 1.34 & 0.22 & 2.41 & 4.21 & 0.70 & 0.576 \\
RCG101 & 1.60 & 0.42 & 5.15 & 6.68 & 0.63 & 0.55\\
RHG116 & 1.65 & 0.29 & 4.98 & 7.50 &1.25 & 0.54 \\
RHG097 & 1.84 & 0.45 & 3.12 & 7.88 & 1.66 & 0.546 \\
RHG036 & 2.22 & 0.20 & 2.78 & 1.96 & 0.7 & 0.571 \\
RHG109 & 2.85 & 0.39 & 2.70 & 6.87 & 1.96 &0.458 \\

\hline
\end{tabular}
\end{flushleft}
\label{tab:paratab}
\end{table}

\begin{figure}
\rotatebox{-90}{\includegraphics[height=8.0 cm]{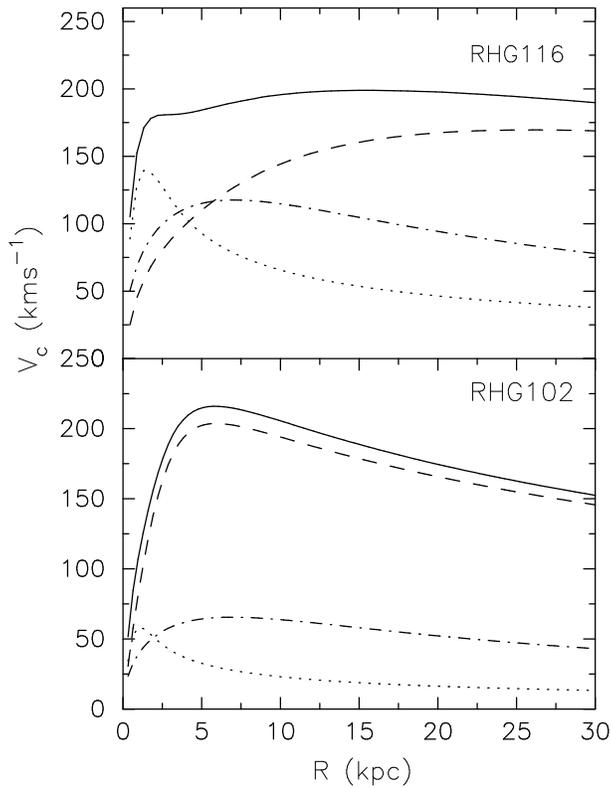}}
\caption{Initial circular velocity curve for the model RHG116 and RHG102. The solid line 
denotes the total circular velocity curve; the dashed line is the contribution 
from dark matter halo, dash-dot-dash line for the disc and dotted line for 
the bulge. The model RHG102 is dark matter dominated throughout.}
\label{fig:vc116}
\end{figure}

We present here a sample of $16$ very thin (with the ratio of scale height to 
scale length $h_z/R_d$ being less than $0.05$ or close to it) equilibrium models 
of disc galaxies composed of a stellar disc, bulge and dark matter halo constructed 
using the method of \cite{KD1995}. Each component in the galaxy model is live and 
interact with each other via gravitational forces.  
The models are scaled so that the unit of scale length ($R_d$) is $4$~kpc and 
the circular velocity at about $2.1 R_d$ is $\sim 200$~kms$^{-1}$. For relevant 
details on model construction, the readers are referred to 
\cite{Sahaetal2010, Sahaetal2012}. The mass of the stellar disc and other physical
parameters of each galaxy model are presented in Table~\ref{tab:paratab}. 
We primarily focus on two fiducial models namely, RHG116 and RHG102, and analyze 
their evolution in detail as a comparative study.
Fig.~\ref{fig:vc116} depicts the initial circular 
velocity curves for RHG116 and RHG102. Both the stellar discs of 
the two models have same initial thickness but the dark matter in the model 
RHG102 is dominating over the disc right from the center whereas dark matter 
contribution dominates in RHG116 only beyond about $2 R_d$.     

\begin{figure}
\rotatebox{-90}{\includegraphics[height=8.0 cm]{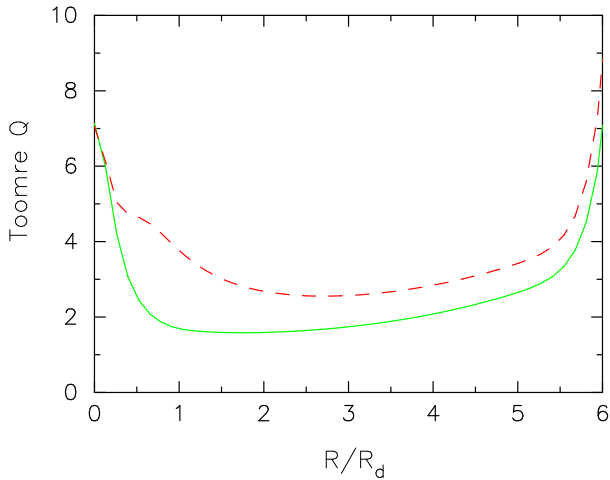}}
\caption{Initial Toomre Q profiles for two models RHG116 (green solid curve) and 
RHG102(red dashed curve).}
\label{fig:tQ}
\end{figure}

Each of the galaxy models was evolved in isolation to understand how an initially 
superthin galaxy would be restructured as result of internal evolution alone. 
The simulations were performed using the Gadget code \citep{Springeletal2001} 
which uses a variant of the leapfrog method for the time integration. 
The gravitational forces between the particles are calculated using the BH tree 
algorithm with a tolerance parameter $\theta_{tol} =0.7$. The integration time 
step used was $\sim 0.4$ Myr and the models were evolved for about $5 - 6$~Gyrs.
The number of particles used for the halo and disk are $1.1 \times 10^{6}$ each, 
the bulge contains $10^{5}$ particles with a total of $2.2 \times 10^{6}$ particles
for each model which is in accordance with the suggestion by \cite{DubinskiBerentzen2009}.  
The softening lengths for the various components were used so that the maximum 
force on each particle is nearly the same \citep{McMillan2007}. The total energy 
is conserved well within 0.2\% till the end of the simulation. 

\section{Internal sources of disc heating}
\label{sec:heating}
In the absence of environmental influences such as tidal interaction, mergers or 
satellite infall, the evolution of superthin galaxies would strongly depend on 
the internal processes. In particular, vertical heating of stars could change the 
thickness of stellar discs and perhaps transform a superthin disc into a typical 
thin or even thick disc. Hence, it is important to identify and understand various 
internal sources that might be operative in heating the stars in superthin galaxies. 

\begin{figure*}
\rotatebox{0}{\includegraphics[height=4.5 cm]{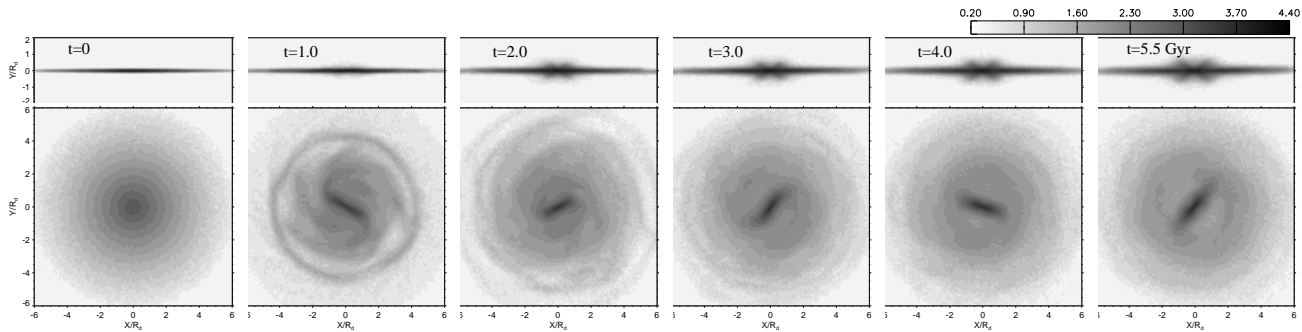}}
\caption{Time evolution of the initially axisymmetric and superthin stellar disc 
in model RHG116. The upper panel shows surface density maps of the stellar disc 
viewed edge-on and the lower panel for the face-on view. Time on each panel 
indicates the time (in unit of Gyr) when each snapshot was taken. Clearly, the disc
makes a transition from superthin to thin.}
\label{fig:den116}
\end{figure*}

Amongst the internal sources of vertical heating the ones that could have been 
important for superthin galaxies are scattering due to giant molecular 
clouds \citep{Jenkinsbinney1990}, substructures or clumpiness in the dark 
matter halo \citep{Fontetal2001, Ardietal2003}, bars and transient spirals 
and/or bending waves \citep{HunterToomre1969,Sahaetal2010}. 
However, the low star formation rates suggest that 
the presence of many giant molecular clouds are unlikely in superthin 
galaxies. Heating due to substructures in the dark halo are shown to be not efficient 
because the masses of the substructures are unlikely to be more than
$10^{11} M_{\odot}$ and the orbits of the substructures rarely passes close 
to the disc mid-plane \citep{Fontetal2001}.  

Damping of bending waves could, in principle, lead to vertical heating 
of stars in galaxies. A detailed study by \cite{Araki1985} showed that a bare 
stellar slab can become prone
to strong bending instability (or firehose instability) if $\sigma_z/\sigma_r < 0.293$.
Some of the stellar discs in our simulation, indeed, have $\sigma_z/\sigma_r < 0.3$ but
we have not noticed any strong bending mode. Possible reason could be that these discs 
have finite thickness and are embedded in spheroidal dark matter halo -- both of which 
provide restoring force (proportional to the square of the total vertical frequency, 
in this context) to stabilize the bending modes \citep{NelsonTremaine1995a,Saha2008,Sellwood2013}. 
In real galactic stellar discs, warps, although common, have small amplitudes
\citep{Sahaetal2009} and are thought to rotate with a slow pattern speed 
(though not measured), see \cite{NelsonTremaine1995a}. As a consequence, an warp
has a little energy (nearly equal to the kinetic energy of a single globular cluster) 
stored in it \citep{NelsonTremaine1995b} to contribute to the vertical heating. 

In N-body simulation, two-body
relaxation due to discreteness noise is a source of heating but the process
is known to be suppressed with increasing the number of particles and a convergent
behaviour is reached once a simulation contains a few million particles 
as suggested by \cite{DubinskiBerentzen2009} and this has been explicitly tested
in a previous paper by \cite{Sahaetal2010}. Note each model galaxy in our sample 
is composed of $2.2 \times 10^6$ particles (also mentioned in the previous section)
in compliance with suggestion by \cite{DubinskiBerentzen2009}. 

\section{Evolution of superthin galaxies}
\label{sec:evolution}
Galaxies are thought to evolve from late-type to early-type along the Hubble 
sequence either through secular evolution or by external influences or a combination 
of the two, in general. It is important to understand the rate at which galaxies 
evolve along such a sequence. The rate at which any late-type galaxies would 
evolve towards early-type would certainly depend on various factors such as star 
formation rate, the efficiency of redistributing energy and angular momentum amongst 
the stars in the host galaxies. As mentioned in section~\ref{sec:intro},
superthin galaxies somehow lack environmental influences. As a consequence, the 
evolution in superthin galaxies are mostly driven by internal mechanisms of which 
bars could be the most likely candidate. It is because galaxies do form bars under 
diverse physical conditions as learned from the studies of N-body simulations 
as well as inferred from the ample abundance of barred galaxies in the local universe. 
Since thickness is one of the striking 
feature which primarily distinguishes superthin galaxies from the rest, we evolve 
models of initially axisymmetric extremely thin galaxies and follow how their 
thickness changes.      

\begin{figure}
\rotatebox{-90}{\includegraphics[height=8.0 cm]{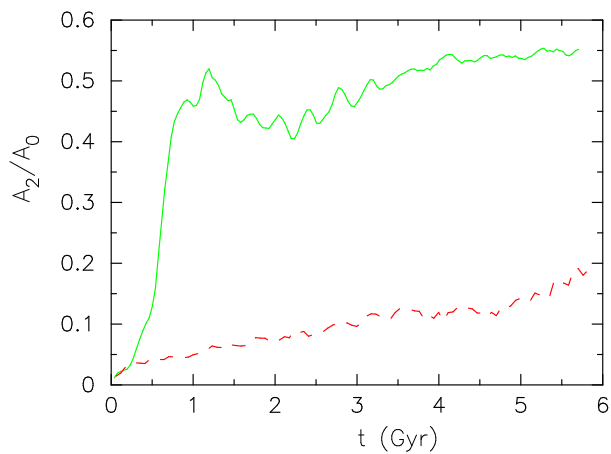}}
\caption{The time evolution of bar strength for models RHG116 (green solid curve) 
and RHG102 (red dashed curve). The model RHG102 develops a weak bar over $\sim 6$~Gyrs.}
\label{fig:A2A0}
\end{figure}

\subsection{Thickness of a stellar disc}
\label{sec:thickness}
During the initial phase of evolution when the disc is still axisymmetric to a 
good approximation, the vertical structure of the disc can be determined assuming 
vertical equilibrium under the total potential. Under the isothermal approximation 
along the vertical direction, the scale height of a stellar disc can then be estimated 
using the relation \citep{BT1987} 

\begin{equation}
h_z (r) = \frac{\sigma_z (r)}{\sqrt{2 \pi G \rho_{mid}(r)}},
\end{equation}

where $\sigma_z$ is the vertical velocity dispersion and $\rho_{mid}$ is the mid-plane 
mass density of stars at a location $r$. The initial density distribution of stars along the vertical 
direction is well approximated using a $sech^2$ law as 
$\rho(r, z)=\rho_{mid} (r)\times sech^2(z/h_z)$ in our model galaxies. However, 
as a galaxy evolves, it forms bars, spiral arms or undergoes bending instability, 
the stars are basically subject to a time-dependent potential. Hence, the above relation for 
determining the scale height of a stellar disc might not hold true any more.

We determine the scale height of a stellar disc at any epoch during the evolution 
using the second moment of the volume density distribution as follows \citep{Sahaetal2009}

\begin{equation}
H(r) = \frac{\int{z^2 \rho(r,z) dz}}{\int{\rho(r,z) dz}}.
\label{eq:Hz}
\end{equation}

In the above equation, $\rho(r,z)$ is the azimuthally averaged density of stars at
a location ($r, z$) in the meridional plane. 
The scale height of the disc is obtained as $H_{z}(r) = \sqrt{H(r)}$. Note that 
for a $sech^2$ density distribution, the method of second 
moment gives a value of the scale height which is $0.907 \times h_z$. The radial 
variation of $H_{z}(r)$ would give the flaring information about the stellar disc. 
Initially, the radial variation of scale height of a stellar disc is flat (by construction) 
and during the subsequent evolution, the stellar discs show mildly flaring. In fact, 
previous work by \cite{Narayan2002a, Narayan2002b} argue that the scale height of 
stars in observed galaxies (such as NGC 891 and NGC 4565) are likely to be moderately 
flaring when one takes into account the gas self-gravity in the hydrostatic balance. 
Moderate flaring of stars are also required in order to explain the onset of warps
in a number of nearby edge-on galaxies \citep{Sahaetal2009a}. In the current work, 
we ignore mild flaring of stars and compute the stellar scale height averaged over 
$3 - 4$ disc scale lengths which is well outside the radial extent of the bar that 
forms in most of our stellar discs and use that value of scale height for 
subsequent analysis. Throughout the text, we use thickness and/or scale height to indicate
the same quantity defined in Eq.~\ref{eq:Hz}. Below we discuss detailed evolution of 
two initially superthin galaxies having same initial thickness that follows two entirely 
different evolutionary path. 

\begin{figure}
\rotatebox{-90}{\includegraphics[height=8.0 cm]{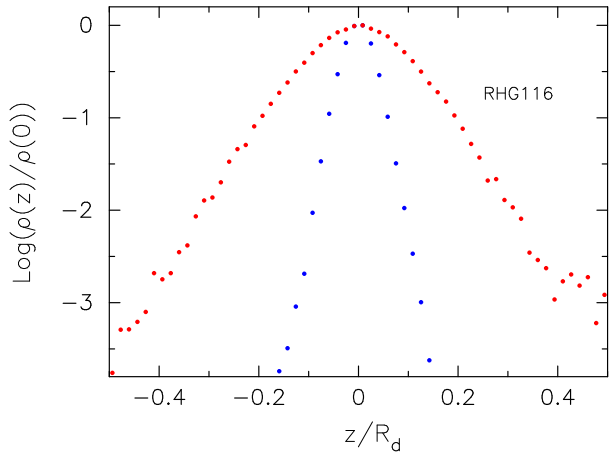}}
\caption{The vertical density distribution of stars for model RHG116. The blue 
dotted curve denotes the initial one and the red dotted curve after $5.5$~Gyrs 
computed at $4$ disc scale lengths. $\rho(0)$ is the midplane density at the 
corresponding epoch.}
\label{fig:rhoz116}
\end{figure}

\begin{figure}
\rotatebox{-90}{\includegraphics[height=8.0 cm]{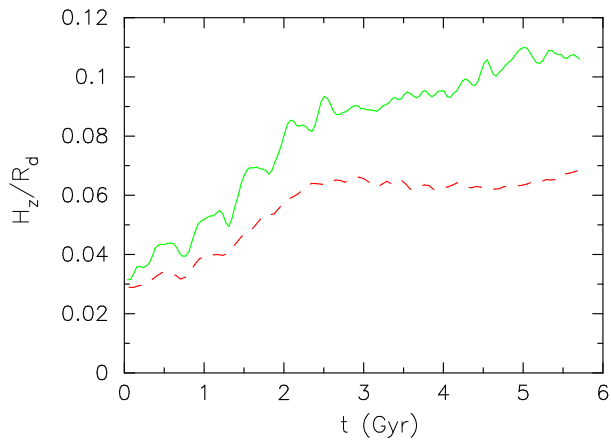}}
\caption{The time evolution of stellar scale height in two models, RHG116 and RHG102.
The green solid curve showing the scale height of stars which are heated by a strong bar 
as in model RHG116. The red dashed line represent the same but for the model RHG102.}
\label{fig:thicktime}
\end{figure}

\subsection{Stellar discs with strong bar}
\label{sec:sbar}
More than $2/3$ of the observed disc galaxies host strong bars in their central 
regions implying bars are common in disc galaxies \citep{Eskridgeetal2000,Barazzaetal2008}. 
However, the theoretical 
understanding of the formation and growth of bars in real galaxies is not 
yet clear \citep{Sellwood2013}. N-body simulations, especially the work 
of \cite{Athanassoula2002} has shown that a bar can grow stronger in 
the presence of a live dark matter halo and recent N-body simulations by a number 
of authors such as \cite{DubinskiBerentzen2009}, 
\cite{Sellwood-Debattista2009}, \cite{Klypinetal2009}, reveal 
further insights on the role of dark matter halo mass distribution in the context 
of bar growth; while \cite{SahaNaab2013} showed the importance of halo spin on the
bar formation. Once formed, a strong bar is shown to be an efficient source of vertical 
heating of the disc stars \citep{Sahaetal2010}. 

The stellar disc of RHG116 is relatively cold with a Toomre $Q =1.65$ (at $2.5 R_d$), 
see Fig.~\ref{fig:tQ} for the radial variation. The disc forms a strong bar within a 
billion year; the evolutionary sequence is depicted through a set of grey-scale images 
(Fig.~\ref{fig:den116}). 
Subsequently, the bar grows in size and undergoes the well-known
buckling instability leading to the formation of a prominent boxy/peanut bulge
\citep{Combesetal1990,Rahaetal1991, HernquistWeinberg1992,Athanamisi2002,
Debattistaetal2004,KormendyKennicut2004,Sahaetal2010,SahaNaab2013} at about 
$2$~Gyr in this model. A corresponding drop in the value of $A_2/A_0$ (here,
$A_m$ denotes the $m^{th}$ order Fourier component of the surface density 
distribution) can be seen from Fig.~\ref{fig:A2A0}. After the buckling phase, the
disc progressively thickens as it is evident from the edge-on images
taken at different epochs during the evolution (see Fig.~\ref{fig:den116}).
During this growing phase of the bar, the stars are subject to chaotic diffusion as shown
by \cite{Brunettietal2011}. In order to find out the impact of this process,
we follow the vertical density distribution of stars at several different locations
in the disc and at several epochs during the evolution.
It is important to notice that the disc contains apparently no spiral arms nor any bending modes
beyond about $2$~Gyr and mostly dominated by a bar which as mentioned above, is capable of stirring
up the stars globally. Although the impact of a bar on the stars are more visible
within the corotation radius,
it is equally important to understand how the vertical distribution of stars
changes outside this radii as a result of bar driven chaotic diffusion which is 
not limited to corotation region. Fig.~\ref{fig:rhoz116} shows 
the initial (t=0) and final (t=5.5~Gyr) vertical density distribution of stars computed at 
$\sim 4 R_d$. This clearly demonstrates how bar alone can fatten the distribution
of stars along the vertical direction and this primarily happens through the
broadening of the velocity distribution function which was previously shown by 
\cite{Sahaetal2010,SahaPfennigerTaam2013}.    

\begin{figure*}
\rotatebox{0}{\includegraphics[height=12.0 cm]{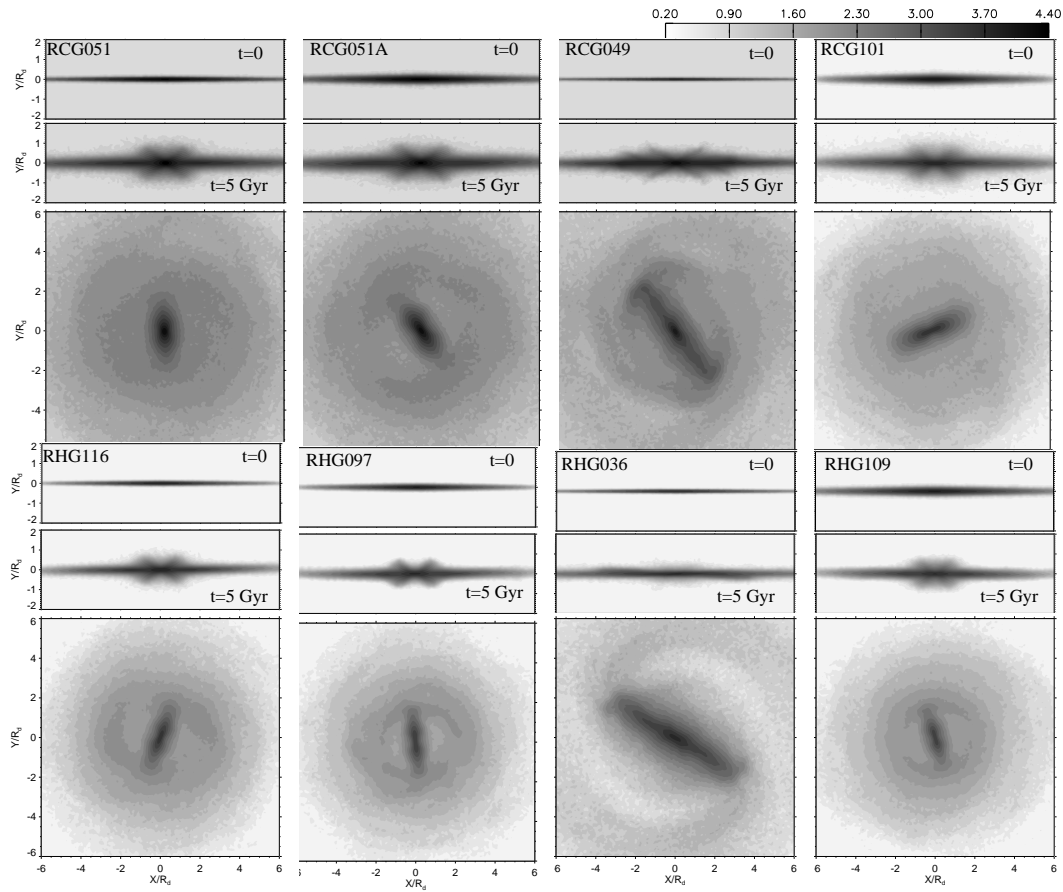}}
\caption{Surface density maps of stellar discs which formed strong bars. For each model galaxy,
we show three such maps: the upper most one at t=0 (edge-on), middle one 
at t=5 Gyr (edge-on) and bottom panel at t=5 Gyr (face-on). All the panel maps are
scaled to same color scale.}
\label{fig:denTGs}
\end{figure*}

In Fig.~\ref{fig:thicktime}, we show the evolution of the stellar scale height 
computed in a way explained in section~\ref{sec:thickness}. 
Our results show that the disc becomes thicker nearly by a factor of $3.7$ in about $5.5$~Gyr. 
In terms of kpc scale, the scale height of stars increases from $120$ pc to $\sim 444$~pc over 
this time period and the stellar disc leaves the superthin regime. Notice that the disc 
does not have any apparent spiral arms associated with the bar nor any strong bending 
oscillations in the outskirts of the disc (see Fig.~\ref{fig:den116}) which becomes 
hot enough for any of these features to survive over such a longer period of time.

We have carried out this analysis for the rest of the models which formed strong bars, see
Fig.~\ref{fig:denTGs}. All of these model galaxies were initially superthin and transformed
to thin ones over about $5$~Gyr. The bar strengths for each model are given in 
Table~\ref{tab:paratab}. An important message is that superthin galaxies are
unlikely to host strong bars along their evolutionary path.
    
\subsection{Stellar discs with weak bar}
Looking at the weak bars (e.g., type SAB), it is not clear whether they are 
formed via partial dissolution of strong bars or due to some other mechanisms 
\citep{SellwoodWilkinson1993}. Understanding the properties of weak bars, 
its formation and growth in real galaxies requires further investigation both 
in observation and numerical studies. \cite{Sahaetal2010} report the formation 
of many weak bars which grow very slowly, typically on a secular 
evolution time scale. From our present study, it appears that weak bars are weak 
from their birth and such weak bars are formed preferentially in galaxies having a higher 
value of Toomre Q (i.e., radially hot) combined with a higher dark matter fraction over the 
stellar counterpart (quantified as $M_h/M_d$), see Table~\ref{tab:paratab}. Details
about one of the weak bar model RHG057 are also presented in \cite{SahaPfennigerTaam2013}.     

\begin{figure*}
\rotatebox{0}{\includegraphics[height=4.5 cm]{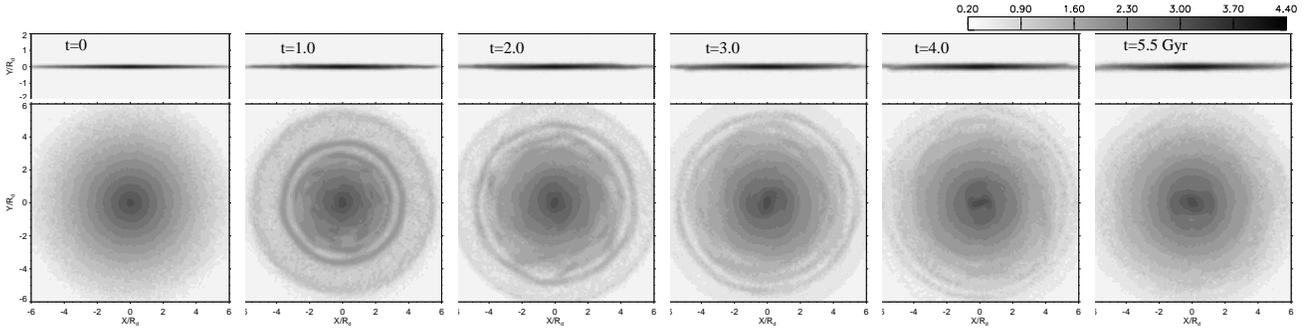}}
\caption{Same as in Fig.~\ref{fig:den116} but for RHG102. Note a very slow
morphological change of the stellar disc compared to that of RHG116.}
\label{fig:den102}
\end{figure*}

\begin{figure*}
\rotatebox{0}{\includegraphics[height=12.0 cm]{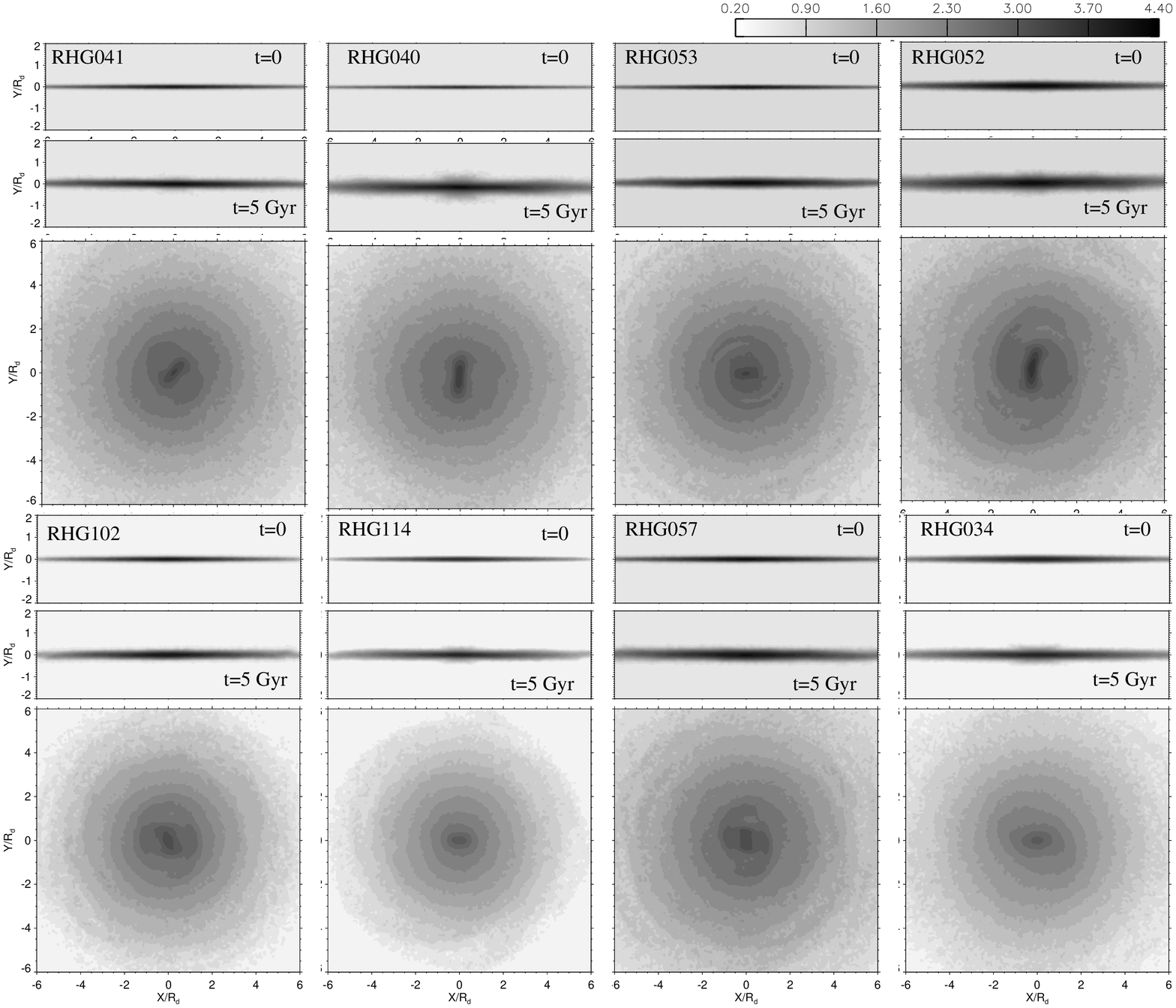}}
\caption{Surface density maps of stellar discs which formed weak bars. For each model galaxy,
we show three such maps: the upper most one at t=0 (edge-on), middle one 
at t=5 Gyr (edge-on) and bottom panel at t=5 Gyr (faceon). All the panel maps are
scaled to same color scale.}
\label{fig:denSGs}
\end{figure*}

The stellar disc of RHG102 is hot with $Q=2.56$ (see Fig.~\ref{fig:tQ} for the radial profile)
and dark matter dominated. Fig.~\ref{fig:den102} shows the surface density maps at
different epochs and the time evolution of the bar amplitude can be seen from
Fig.~\ref{fig:A2A0}. What we learn from these figures is that
the growth of the bar in this model is very slow and remains weak till $\sim 6$~Gyr which 
is nearly half the Hubble time. Some of these models were run for even longer time e.g.,
RHG057 which did not grow any stronger bar \citep[see,][]{SahaPfennigerTaam2013}. 
The edge-on view of the stellar disc in RHG102 shows no obvious 
sign of buckling instability because the bar is not self-gravitating enough and the disc 
lacks vertical inner Lindblad resonance (ILR) \citep{Rahaetal1991,PfennigerFriedli1991}. 
From the face-on view of this model galaxy, it is clear that this 
model also has not formed any strong two-armed spiral and the edge-on view 
reveals no obvious bending oscillations either; altogether the stellar disc 
is able to maintain a smooth, featureless structure over a significant fraction of
the Hubble time. This holds true for the rest of the model galaxies in our sample
(see, Fig.~\ref{fig:denSGs}) which grow nothing but weak bars over a longer period of time.

\begin{figure}
\rotatebox{-90}{\includegraphics[height=8.0 cm]{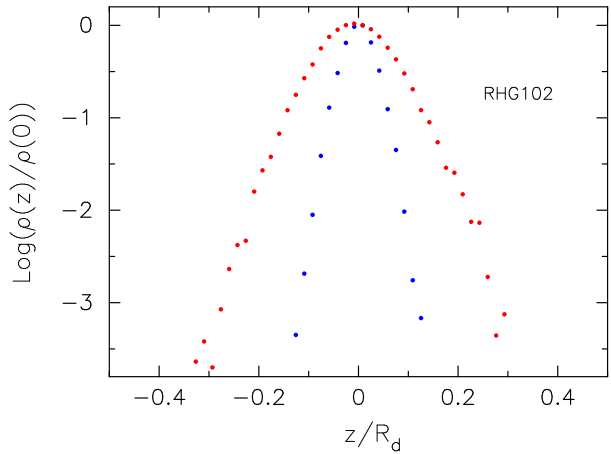}}
\caption{Same as in Fig.~\ref{fig:rhoz116} but for the model RHG102.}
\label{fig:rhoz102}
\end{figure}

The time evolution of the stellar disc thickness in model RHG102 can be seen from
Fig.~\ref{fig:thicktime}. The thickness increases roughly by a factor 
of $2$ in $2.5$~Gyr and remained nearly unchanged till about $6$~Gyr - 
indicating a very slow heating process. In terms of kpc scale, the final scale 
height of stars at $5.5$~Gyr is $\sim 256$ pc indicating that the stellar disc
can be considered as one of superthin category. Note that the stellar disc of
our Milky Way has a scale-height of about $400$~pc. The vertical density distribution 
of stars at $5.5$~Gyr resembles quite well the initial $sech^2$ profile 
(see Fig.~\ref{fig:rhoz102}). In other words, the vertical density near the mid-plane
changes quantitatively by a very small amount. It turns out that a plausible evolutionary 
scenario for the 
superthin galaxies is that they grow weak bars in the central region and
are evolving on a very slow secular evolution time scale. One could, in principle, argue
that the observed superthin galaxies might not be forming even weak bars; but the
point we are making here is that it is hard for a galaxy to remain completely
axisymmetric over a Hubble time as demonstrated for a range of models in Fig.~\ref{fig:denSGs}.    

\subsection{Dependence of disc thickness on bar strength and dark matter haloes}
\label{sec:depend}
\begin{figure}
\rotatebox{-90}{\includegraphics[height=8.5 cm]{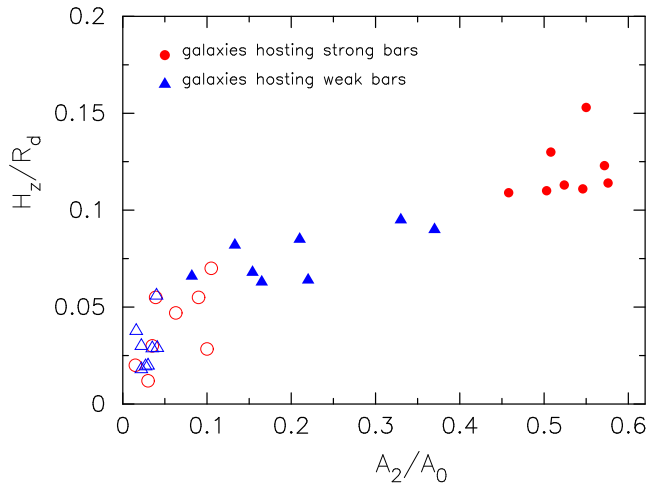}}
\caption{Galaxy evolution in the parameter space spanned by bar strength 
and stellar thickness. Initial $A_2$'s are measured within $T_{orb}$. The red 
open circles (initial models) are evolved into filled red circles and blue open 
triangles (initial models) are evolved into filled blue triangles in $5 - 6$~Gyrs.}
\label{fig:A2vshz}
\end{figure}

Here, we investigate how these galaxies are distributed in the 
parameter space spanned by the thickness of stars and bar strength. We carry out 
this exercise at different epochs over a period of $5 -6$~Gyr. The initial and final 
disc thickness are computed using the method described in section~\ref{sec:thickness}. 
We calculate the initial bar strength (considered as the peak of $A_2/A_0$) when
the disc has been evolved by an orbital time (which is $\sim 300$~Myr). Here
orbital time is measured at the disc half-mass radius. The final bar strength
is calculated at around $5 - 6$~Gyr. Fig.~\ref{fig:A2vshz} shows the 
initial and final bar strengths and their corresponding stellar thickness measurements.
Initially, all the model galaxies are essentially superthin i.e., $H_z/R_d < 0.1$. 
Galaxies that formed strong bars leave the superthin regime in 
$5 - 6$~Gyr time scale transforming into typical thin discs. In contrast, galaxies
that host weak bars remained superthin even after $5 - 6$~Gyr of evolution. 
There seems to be a continuous well defined trend (see Fig.~\ref{fig:A2vshz}) 
in which the thickness of a galaxy's stellar disc varies with the strength
of a bar. In no cases, we found galaxies hosting strong bars maintaining its 
superthinness over a long period of time neither found a case where a galaxy 
hosting a weak bar and having very fat/thick disc in our sample of galaxies 
studied here. This has clearly demonstrated that the galaxies undergoing strong bar 
instability can not remain superthin over several billion years. In other 
words, if at all, superthin galaxies that we see today would be hosting 
only weak bars.

\begin{figure}
\rotatebox{-90}{\includegraphics[height=8.5 cm]{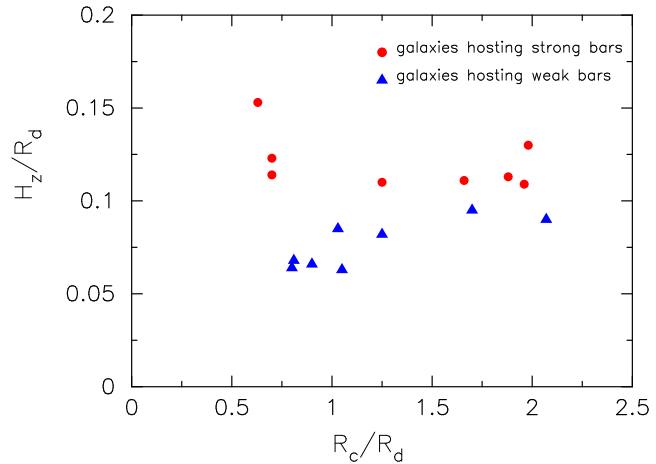}}
\caption{Galaxy evolution in the parameter space spanned by core radii of dark halos and 
stellar scale height. All model galaxies are evolved for about $5 - 6$~Gyrs.}
\label{fig:Rcvshz}
\end{figure}

The strength and long-term evolution of a bar are intricately related to the dark 
matter distribution in the host galaxy, especially in the central region 
\citep{Athanassoula2003, DubinskiBerentzen2009}. In this subsection, we examine 
whether the growth and final strength of a bar (which eventually thicken the 
stellar disc) depends explicitly on the core radius ($R_c$) of the surrounding 
dark matter halo. In Fig.~\ref{fig:Rcvshz}, we show the thickness of all the stellar discs and 
the corresponding $R_c/R_d$ values of the dark matter haloes calculated at the 
end of $5-6$~Gyr. For our galaxy models, the values of $R_c/R_d$ vary from as low
as $\sim 0.5$ to about $2.1$. What we find is that the dark matter haloes of 
superthin galaxies can have a wide range of halo core radius; the same applies to
galaxies that are classified as thin ones in their final state. It appears
that the galaxies in our sample that maintained their superthinness over 
$5 -6$~Gyr showed no strong preference for any particular dark matter haloes.    

By studying the hydrostatic equilibrium of stars in the presence of gas and dark matter
halo, \cite{BanerjeeJog2013} suggest that the dark matter halo in UGC 7321 is
compact with $R_c/R_d =1.38$. This is in accordance with what we have found i.e., all 
of our final superthin galaxies have $R_c/R_d < 2$ except one with $R_c/R_d \simeq 2.1$. 
What is intriguing is that even thin/thicker galaxies could be surrounded by such
compact haloes.
It is worth mentioning that UGC 7321 is an well-studied case of superthin galaxies
\citep{Matthewsetal1999,Matthews2000}. Their study suggests that the superthin 
galaxy UGC 7321 is an underevolved system both in dynamical and star forming sense. 
From the disc colors and color gradients, they also indicate that UGC 7321 is not 
a young galaxy. There is a population of old disc stars above the disc midplane 
indicating some ongoing dynamical heating in the galaxy. Our study suggests that 
weak bars could be responsible for a slow heating of the stars in the galaxy. In fact, 
the position velocity diagram in HI line \citep{Uson2003} and the deviation of 
the light profiles in the inner region, the shape of the isophotes in 
R-band \citep{Pohlenetal2003} --all together indicate that there might be a thin
bar hidden in the galaxy, UGC 7321. Based on our present analysis, we can comment that 
such a thin bar (if present in UGC 7321) must be weak; otherwise 
it would have gone through a bar buckling instability transforming the superthin disc to
a thin one. 

\section{discussion and conclusions}
\label{sec:discuss}

The very presence of smooth, featureless superthin galaxies in our local 
universe suggest that they are not subject to strong minor mergers or 
significant accretion events, as otherwise they would either be destroyed 
or converted to thicker disc galaxies \citep{Toth1992, Purcelletal2009}. 
And if they are isolated as observation indicates, they should be subject 
to disc heating arising due to the internal sources mentioned 
in section~\ref{sec:heating}. Obviously, if we could switch off some of these
heating sources, the issue of maintainance of superthinness could be resolved.
Our simulations suggest that this is unlikely as stellar discs, with diverse
initial condition, form either strong or weak bars both of which heat stars
but of course, with varying efficiency.
On the other hand, the apparently smooth, flat stellar discs also indicate that SGs are unlikely 
to host strong bars as otherwise such discs eventually would have to face 
a buckling instability \citep[see][for coherent review]{Sellwood2013}. 
The other possibility, which can not be ruled out unambiguously, 
is that SGs do form weak bars over a longer time scale (as shown by our simulations).

The question arises on the validity of the initial condition that we assume.
Were progenitors of present day superthin galaxies really radially hot?
If yes, how those progenitors achieved such a radially hot discs?
The answer, of course, remains unknown as it depends on how these galaxies
were actually formed. We speculate that during the early phase of galaxy formation, an 
extremely thin disc would have gone through strong spiral instabilities and 
the amount of preferential radial heating produced by it, would have self-destroyed the 
spiral arms leaving a red-hot dead disc.
The low star formation rates as suggested by \cite{Matthewsetal1999}, indicates 
that the present day superthin galaxies are radially hot preventing them from 
forming further strong non-axisymmetric instabilities.  
Then our study shows that during the course of secular evolution such radially 
hot stellar discs when embedded in a massive live dark matter halo form only 
weak bars. {\it In other words, our simulations suggest 
that weak bars are perhaps the maximal non-axisymmetric features that a self-consistent
superthin galaxy might be able to support if left isolated for several billion years.}

Our main conclusions from the present work are the following:

1. We show that an initially thin stellar disc is able to maintain its thinness
over several billion years if it hosts a weak bar. Such weak bars
heat the stars very slowly and can increase the stellar scale-height roughly 
by a factor of $2$ in about $5 - 6$~ billion years. 
 
2. Our simulations suggest that superthin discs with strong bars are
unlikely as well as weak bars making thicker discs. There seems to be
a good correlation between the thickness of a stellar disc and the
amplitude of the bar it hosts.

3. Our results show that there is no strong preference for smaller 
halo core radii amongst superthin galaxies. Thicker discs could also
reside in halos with smaller core radii. However, our study do not
cover a wide range of halo core radius to disc scale length ratios.

4. We show that during the course of evolution, the underlying nature of the 
vertical density profile in a model hosting a weak bar remains unchanged 
except fattening by a small amount.
  
\section*{Acknowledgement}
\noindent The author acknowledges the support by the Geneva Observatory where
a part of this work was completed.

\end{document}